\documentclass[graybox]{svmult}
\usepackage{mathptmx}
\usepackage{helvet}
\usepackage{courier}
\usepackage{type1cm}
\usepackage{makeidx}
\usepackage{graphicx}
\usepackage{multicol}
\usepackage[bottom]{footmisc}
\def\gtsim{\mathrel{\hbox{\rlap{\hbox{\lower4pt\hbox{$\sim$}}}\hbox{$>$}}}}
\def\ltsim{\mathrel{\hbox{\rlap{\hbox{\lower4pt\hbox{$\sim$}}}\hbox{$<$}}}}
\makeindex
\begin{document}
\title*{On the Fractal Distribution of HII Regions in Disk Galaxies}
\titlerunning{Fractal Distribution of HII Regions}
\author{N\'estor S\'anchez and Emilio J. Alfaro}
\authorrunning{N. S\'anchez and E. J. Alfaro}
\institute{N\'estor S\'anchez \& Emilio J. Alfaro \at Instituto
de Astrof\'{\i}sica de Andaluc\'{\i}a, CSIC, Granada, Spain.
\email{nestor@iaa.es, emilio@iaa.es}}
\maketitle
\abstract{In this work we quantify the degree to which star-forming
events are clumped. We apply a precise and accurate technique to
calculate the correlation dimension $D_c$ of the distribution of
HII regions in a sample of disk galaxies. Our reliable results
are distributed in the range $1.5 \ltsim D_c \ltsim 2.0$. We
get significant variations in the fractal dimension among
galaxies, contrary to a universal picture sometimes claimed
in literature. The faintest galaxies tend to distribute their
HII regions in more clustered (less uniform) patterns. Moreover,
the fractal dimension for the brightest HII regions within
the same galaxy seems to be smaller than for the faintest
ones suggesting some kind of evolutionary effect.}

\section{Introduction}

There is clear evidence that gas observed in external galaxies
(both irregulars and spirals) follows hierarchical and self-similar
patterns \cite{Wes99,Kim03,Wil05,Beg06,Dut08}. This
fractal picture is consistent with the distribution of star
fields and star-forming sites on galaxy-wide scales
\cite{Fei87,Par03,Elm06,Fue06,Ode06,Bas07}. However, it is
not clear whether this kind of fractal distributions are
connected/related or not to some properties of the host
galaxies, such as radius, rotation, brightness, morphology, etc.
In spite of the great variety of fractal dimension values reported
in the literature for different galaxies (for both the gas
and the distribution of star forming sites) most of the
authors argue in favor of a more or less universal picture.
In this universal description, the constancy of the fractal
dimension is a natural consequence of the fact that the same
physical processes are structuring these systems. However,
there are some indications that the situation could be more
complicated. The fractal
dimension of the distribution of HII regions could be different
in grand design and flocculent galaxies \cite{Hod85}, and the
brightest galaxies could have fractal dimensions higher than
faintest ones \cite{Par03,Ode06}. On the contrary, \cite{Fei87}
do not find any correlation between the fractal dimension and
the galactic properties, but their uncertainties are so large
that the robustness of this conclusion is questionable.

Part of the problem that prevents achieving unequivocal
conclusions lies in the great diversity of analysis
techniques used in the literature and/or the application
to not large enough samples of galaxies. Here we use a
carefully designed method that has been tested previously
on simulated data and that clearly establishs its accuracy
and applicability depending on the sample itself. We apply
this method to the most complete sample of galaxies that
we have found in literature expecting to draw significant
conclusions regarding this matter.

\section{Fractal dimension calculation}

We use the correlation dimension \cite{Gra83} which gives
robust results when dealing with distributions of points
in space. First we have to calculate correlation integral
$C(r)$ which represents the probability of finding a
point within a sphere of radius $r$ centered on another
point. For a fractal set $C(r)$ scales at small $r$ as
\begin{equation}
\label{eq:dc}
C(r) \sim r^{D_c} \ \ ,
\end{equation}
being $D_c$ the correlation dimension. When evaluating $C(r)$
for real data this power-law behavior is valid
only within a limited range of $r$ values. For
relatively small or high $r$ values $C(r)$ deviates
from the expected behavior and $D_c$ tends to be
underestimated. We have developed an algorithm
(see details in \cite{San07GB} and \cite{San08}) to
calculate $D_c$ which implements
objective and suitable criteria to avoid both
boundary effects and finite-data problems at small scales.
Moreover, the algorithm estimates an uncertainty associated
to $D_c$ by using bootstrap techniques ($\sigma_{boot}$).

\begin{figure}[t]
\sidecaption[b]
\includegraphics[scale=.55]{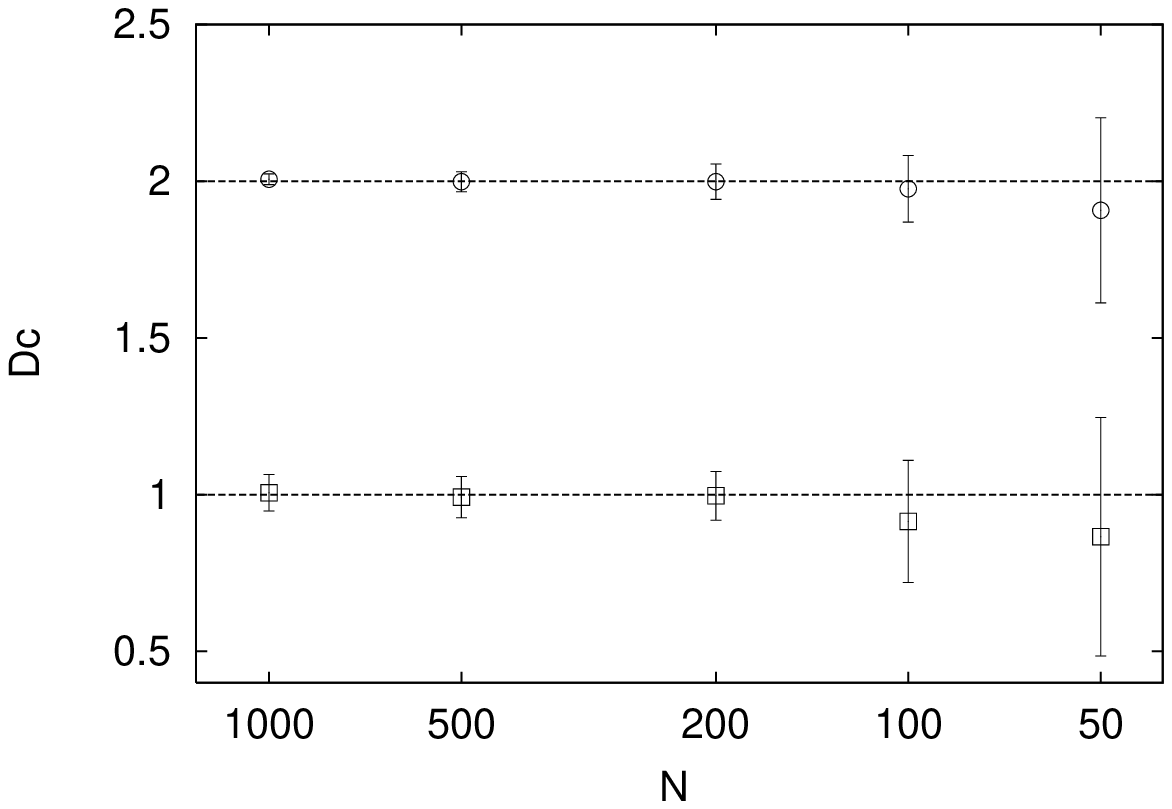}
\caption{Calculated dimension $D_c$ as a function of the
sample size $N$ for fractals with expected dimensions
$D_c=2$ (open circles) and $D_c=1$ (open squares).
Each point is the average of $50$ random realizations
and the bars show the standard deviations.\vspace{1cm}}
\label{fig1}
\end{figure}

An important point is the dependence of the measured
dimension on the sample size because many times the number
of available data points is rather small. Figure~\ref{fig1}
shows what happens when the calculation is done on the
{\it same set of simulated fractals} but randomly removing
points to reach smaller sample sizes $N$. Since we are
dealing exactly with the same original data, the observed
decreasing in the average $D_c$ value with decreasing $N$
have to be attributed exclusively to sample size itself.
Thus, our algorithm is able to estimate the fractal
dimension in a reliable way when applied to random
subsamples of fractal distributions of points, but if
the sample size is too small ($N \ltsim 100$) a bias
tending to underestimate the dimension is produced.
The existence of this bias is related to the random
variations in the simulated (or observed) data set and
not related to the uncertainties in the determination of
$D_c$ ($\sigma_{boot}$). The latter ones are always smaller
than the error bars in Figure~\ref{fig1}.

\section{Results}

We have used VizieR and ADS databases, in conjunction
with the papers \cite{Gar91} and \cite{Gar02}, to search
for catalogs of external galaxies
containing positions of HII regions available
in machine readable format. We found a total of 93
spiral galaxies with positions for at least 50 HII
regions. We have also included data for 8 irregular
galaxies with HII positions provided by D. Hunter
\cite{Roy00}. The properties of the selected galaxies
are listed in Table~2 of \cite{San08} available through
VizieR at CDS. This table contains the galaxy name,
Hubble type, position $\phi$ and inclination $i$
angles, morphological de Vaucouleurs type $T$, arm class
$A$, distance $D$, B-band absolute magnitude $M_B$,
radius of the isophotal 25 mag/arcsec$^2$ $R_{25}$,
and the maximum rotation velocity $V_{rot}$.
The deprojected positions of the HII regions
for the sample of 101 disk galaxies, which can be seen
in Figure~3 of \cite{San08}, are also available through
VizieR.

\begin{figure}[t]
\includegraphics[scale=1.0]{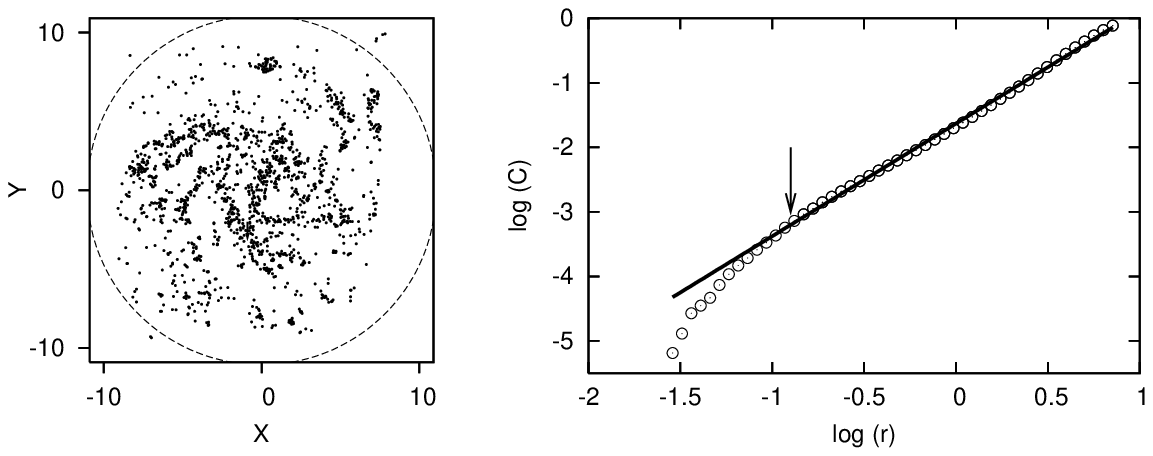}
\caption{Left panel: positions (units are in kpc) of the HII
regions in NGC~6946. The dashed line circle indicates the
radius $R_{25}$. Right panel: the calculated correlation
integral $C(r)$ for NGC~6946. The lower limit of the best
linear fit (solid line) is indicated by the vertical arrow.}
\label{fig2}
\end{figure}

Figure~\ref{fig2}
shows the result for one example galaxy. The left
side panel in this figure shows the distribution of
HII regions in NGC~6946. The
right side panel shows the corresponding correlation
integral $C(r)$. The vertical arrow indicates the value
of $r$ below which $C(r)$ is poorly estimated
\cite{San07GB,San08}. The rest of the data exhibits
a characteristic fractal behavior. The slope of the
best linear fit gives $D_c$, in this case $D_c=1.75$.

\begin{figure}[t]
\sidecaption[b]
\includegraphics[scale=.55]{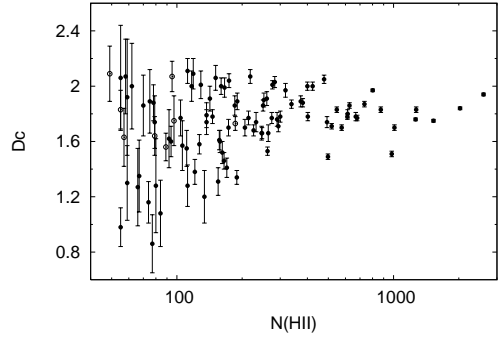}
\caption{Calculated dimension $D_c$ as a function of the
number of available data $N(\mathrm{HII})$ for spiral
galaxies (solid circles) and irregular galaxies
(open circles) in the sample. The error bars indicate
the uncertainties obtained from bootstrapping.\vspace{1cm}}
\label{fig3}
\end{figure}

The result of calculating the correlation dimension of
the distribution of HII regions in the full sample of galaxies
is shown in Table~3 of \cite{San08} (also available via
VizieR). Figure~\ref{fig3}
shows the calculated correlation dimension for all the
galaxies in the sample as a function of the number of
available data points.
We see that as $N(\mathrm{HII})$ decreases, the
uncertainties increase and the obtained dimensions
are substantially more spread out toward lower values.
This trend is very similar to the behavior shown in
Figure~\ref{fig1}. It is clear that at least part of
this trend is due to a bias in the estimated
value of $D_c$ for galaxies with small number of HII
regions. To overcome this bias, we focus our
analysis on galaxies having more than 200
HII regions (46 spiral galaxies). The average fractal
dimension in this case is $\langle D_c \rangle = 1.81$
with a standard deviations of $\sigma = 0.14$, and
the average $\sigma_{boot}$ is $\sim 0.03$.
By considering the associated uncertainties
we can say that the $D_c$ values do differ among
themselves.

\subsection{Dependence on galactic properties}

To examine possible correlations we propose, in a first
approximation, a linear model linking all variables and we
evaluate its ``goodness" via the Akaike's Information Criterion
\cite{Aka74}. This criterion not only rewards goodness of fit,
but also includes a penalty that is an increasing function of
the number of free parameters to be estimated (discouraging
overfitting). Then, starting from a linear regression model
based on $n$ variables we apply a stepwise regression process
to select the subset of variables providing the best model.
We begin with the variables $T$, $A$, $M_B$, $R_{25}$,
$V_{rot}$, and another variable that we have called the
average surface density of star forming regions
($N(\mathrm{HII})/R_{25}^2$). The application of
this procedure using the {\bf R} environment
for statistical computing \cite{R08} yields that $M_{B}$
and $R_{25}$ represent the best linear model fitting
the data:
\begin{equation}
\label{eq:fit}
D_c= -(0.069 \pm 0.025) M_{B} -(0.006 \pm 0.003) R_{25} +0.456 \ \ .
\end{equation}
The overall fit is significant at the confidence level of 95\% and
the regression coefficients of the variables $M_{B}$ and $R_{25}$
are significant at the confidence levels of 99\% and 95\%, respectively.

According to our results $D_c$ correlates with $M_{B}$ and, to a
lesser extent, with $R_{25}$. Brightest galaxies have fractal
dimensions higher than faintest ones, while \cite{Fei87} find a pure
scatter diagram when comparing $D_c$ and $M_B$ which, together
with the lack of correlation with other properties is used as
argument favoring a universal $D_c$ value in spiral galaxies.
In contrast, \cite{Par03} and \cite{Ode06} report the same
trend in their samples of dwarf galaxies, i.e. the highest
fractal dimensions for the brightest galaxies.

\subsection{Possible evolutionary effects}

\begin{figure}[t]
\includegraphics[scale=.65]{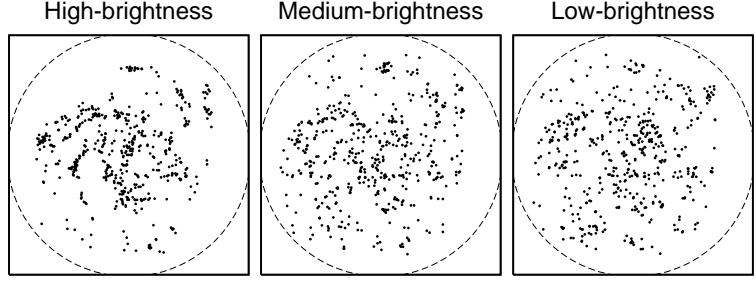}
\caption{Positions of the HII regions in NGC~6946 for high,
medium, and low brightness regions, for which the
resulting fractal dimensions were $D_c = 1.64$, $D_c = 1.82$,
and $D_c = 1.79$, respectively.}
\label{fig4}
\end{figure}

We have analyzed the dependence of $D_c$ on the brightness
of the HII regions. We first selected galaxies from the
sample for which we have data on HII region brightness and
a sample size of $N(\mathrm{HII}) > 600$. There are 9 galaxies
in our sample fulfilling these requiriments. We divided them in 
three equal subsamples ordered in descending brightness:
the brightest HII regions, the medium bright ones, and
the faintest ones. Figure~\ref{fig4}
shows the resulting distributions
in the galaxy NGC~6946. It can be seen that the obtained $D_c$
values are consistent with the appearance of the HII region
distributions. The high brightness map has a relatively small
$D_c$ value ($\simeq 1.6$) that corresponds to a more irregular
distribution having clumps and/or filaments separated by low density
(or empty) regions. The medium and low brightness maps have higher
$D_c$ values ($\simeq 1.8$) corresponding to more homogeneous
distributions. We recalculated $D_c$ for each case and
found that for 7 of the 9 galaxies $D_c$ is smaller for the
brightest HII regions than for the rest of the data.
The brightest regions should reflect, in a first approximation,
the initial distribution of star-forming regions in each galaxy.
It seems likely from our results that some kind of evolutionary
effect tends to randomize (homogenize) in some degree the initial
distributions of HII regions.

\section{Final remarks}

There are many other galactic properties from which $D_c$
could depend, such as star formation activity, mass,
age, metallicity, or a combination of them. It has been
suggested that the fractal
dimension of the distribution of star-forming sites could be
increased during the star formation process \cite{Fue06}.
Recently, a transition scale from a lower to a higher
correlation dimension has been suggested as a possible
explanation for the wide range of observed $D_c$ values
\cite{Ode08}. Obviously, a more complete analysis including
more galactic variables and a wider and more diverse sample
of galaxies would be necessary in order to obtain a clearer
picture.

\begin{acknowledgement}
We acknowledge financial support from MEC of Spain through grant
AYA2007-64052 and from Consejer\'{\i}a de Educaci\'on y Ciencia
(Junta de Andaluc\'{\i}a) through TIC-101.
\end{acknowledgement}


\begin{thebibliography}{99.}
\bibitem{Aka74} Akaike, H. 1974,
         IEEE Transactions on Automatic Control, 19, 716
\bibitem{Bas07} Bastian, N., Ercolano, B., Gieles,
         M., Rosolowsky, E., Scheepmaker, R.~A., Gutermuth, R., \&
         Efremov, Y. 2007, MNRAS, 379, 1302 
\bibitem{Beg06} Begun, A., Chengalur, J. N., \& Bhardwaj, S. 2006,
         MNRAS, 372, L33
\bibitem{Fue06} de la Fuente Marcos, R., \& de la Fuente Marcos, C.
         2006, MNRAS, 372, 279 
\bibitem{Dut08} Dutta, P., Begum, A., Bharadwaj, S.,
         \& Chengalur, J. N. 2008, MNRAS, 384, L34
\bibitem{Elm06} Elmegreen, B. G., Elmegreen, D.  M., Chandar, R.,
         Whitmore, B., \& Regan, M. 2006, ApJ, 644, 879 
\bibitem{Fei87} Feitzinger, J. V., \& Galinski, T. 1987, A\&A, 179, 249
\bibitem{Gar91} Garcia-Gomez, C., \& Athanassoula, E. 1991, A\&AS, 89, 159
\bibitem{Gar02} Garcia-Gomez, C., Athanassoula, E., \& Barbera, C. 2002,
         A\&A, 389, 68
\bibitem{Gra83} Grassberger, P., \& Procaccia, I. 1983, Phys. Rev. Lett.,
         50, 346
\bibitem{Hod85} Hodge, P. 1985, PASP, 97, 688 
\bibitem{Kim03} Kim, S., Staveley-Smith, L., Dopita, M. A., Sault, R. J.,
         Freeman, K. C., Lee, Y., \& Chu, Y.-H. 2003, ApJS, 148, 473
\bibitem{Ode06} Odekon, M. C. 2006, AJ, 132, 1834
\bibitem{Ode08} Odekon, M. C. 2008, ApJ, 681, 1248
\bibitem{Par03} Parodi, B. R., \& Binggeli, B. 2003, A\&A, 398, 501
\bibitem{R08} R Development Core Team 2008,
         {\it R: A Language and Environment for Statistical Computing},
         R Foundation for Statistical Computing, Vienna, Austria,
         URL http://www.R-project.org, ISBN 3-900051-07-0
\bibitem{Roy00} Roye, E. W., \& Hunter, D. A. 2000, AJ, 119, 1145
\bibitem{San07GB} S\'anchez, N., Alfaro, E. J., Elias, F., Delgado, A. J., \&
         Cabrera-Ca\~no, J. 2007, ApJ, 667, 213
\bibitem{San08} S\'anchez, N. \& Alfaro, E. J. 2008, ApJS, 178, 1
\bibitem{Wes99} Westpfahl, D. J., Coleman, P. H., Alexander, J., \& Tongue,
         T. 1999, AJ, 117, 868
\bibitem{Wil05} Willett, K. W., Elmegreen, B. G., \& Hunter, D. A.\ 2005,
         AJ, 129, 2186
\end{thebibliography}
\end{document}